\documentstyle[12pt,epsfig,a4,times]{article}
\makeatletter
\def\lessim{\mathrel{\mathpalette\vereq<}}
\def\vereq#1#2{\lower3pt\vbox{\baselineskip1.5pt \lineskip1.5pt
\ialign{$\m@th#1\hfill##\hfil$\crcr#2\crcr\sim\crcr}}}
\makeatother
\begin{document}
\twocolumn[\noindent\begin{center}
\Huge
\bfseries{Stellar Production Rates of Carbon and Its Abundance in the Universe}
\vspace*{1.3ex}

\noindent
{\large \bf H.~Oberhummer$^{1*}$, A.~Cs\'ot\'o$^2$,
H.~Schlattl$^3$}
\vspace*{2ex}

\noindent
\footnotesize
$^1$ Institute of Nuclear Physics, Vienna University of Technology,
Wiedner Hauptstrasse 8-10, A-1040 Vienna,
Austria\\
$^2$ Department of Atomic Physics, E\"otv\"os University,
P\'azm\'any P\'eter s\'et\'any 1/A, H-1117 Budapest, Hungary\\
$^3$ Max-Planck-Institut f\"ur Astrophysik,
Karl-Schwarzschild-Str.~1, D-85741 Gar\-ching, Germany\\
$^*$ To whom correspondence should be addressed. E-mail:
ohu@kph.tuwien.ac.at
\end{center}
\vspace*{5ex}

\small
\noindent
{\normalsize \bfseries{Abstract:}}
The bulk of the carbon in our universe is produced in the triple-alpha
process in helium-burning red giant stars.
We calculated the change of the triple-alpha reaction rate in a microscopic
12-nucleon model of the $^{\bf 12}$C nucleus and looked for the effects of
minimal variations of the
strengths of the underlying interactions. Stellar
model calculations were performed with the alternative reaction rates.
Here, we show that outside a narrow window of 0.5 and 4\,\% of the
values of the strong and Coulomb forces, respectively, the stellar production
of carbon or oxygen is reduced by factors of 30 to 1000.\vspace*{5ex}
]

\noindent
\normalsize
The formation of $^{12}$C through the triple-alpha process takes place in two
sequential steps in the He-burning phase of red giants. In the first
step, the
unstable $^8$Be with a lifetime of only about $10^{-16}$\,s is formed in a
reaction equilibrium with the two alpha particles, $\alpha + \alpha
\rightleftharpoons {^8{\rm Be}}$. In the second step, an additional alpha
particle is captured, $^8{\rm Be}(\alpha,\gamma){^{12}{\rm C}}$.
Without a suitable resonance in $^{12}$C, the triple-alpha rate would be
much too small to account for the $^{12}$C abundance in our
universe. Hoyle~{\it(1)\/} suggested that a resonance level in $^{12}$C, at
about 300-400\,keV above
the three-alpha threshold, would enhance the triple-alpha reaction rate and
would
explain the abundance of $^{12}$C in our universe. Such a level was
subsequently found experimentally when a resonance that possessed
the required properties was discovered {\it (2, 3)\/}. It is the second 0$^+$
state
in $^{12}$C, denoted by $0_2^+$.
Its modern parameters {\it (4)\/} are $\varepsilon = (379.47 \pm 0.18)$\,keV,
$\Gamma = (8.3 \pm 1)$\,eV, and $\Gamma_{\gamma} = (3.7 \pm 0.5)$\,meV, where
$\varepsilon$ is the resonance energy in the center-of-mass frame relative
to the three-alpha threshold, and $\Gamma$ and $\Gamma_\gamma$ are the
total width and radiative width, respectively.

The isotope $^{12}$C is synthesized further in the He burning in
red giants by alpha capture to the O isotope $^{16}$O, leading
to an abundance ratio in the universe of $^{12}{\rm C}:{^{16}{\rm O}}
\approx 1:2$ {\it (5)\/}. If the carbon abundance in the universe were
suppressed by
orders of magnitude, no carbon-based life could have developed in the universe.
But the production of O is also necessary because no spontaneous
development of carbon-based life is possible without the existence of water.

Here, we investigated the abundance ratios of C and O by
starting from slight variations of the strength of the nucleon-nucleon (N-N)
interaction with a microscopic 12-nucleon model. In previous studies, only
hypothetical ad hoc shifts of the resonance energy of the $0_2^+$
state were considered {\it (6)\/}. Some preliminary results of our
calculations are reported elsewhere {\it (7)\/}.

The resonant reaction rate for the triple-alpha process ($r_{3\alpha}$)
proceeding via the
ground state of $^8$Be and the $0^+_2$ resonance in $^{12}$C is given
approximately by {\it (5)\/}
\begin{equation}
\label{alphaa}
{\textstyle r_{3\alpha} = 3^{3/2} N_{\alpha}^3
\left(\frac{2 \pi \hbar^2}{M_{\alpha} k_{\rm B} T}\right)^3
\frac{\Gamma_{\gamma}}{\hbar} \exp \left(- \frac{\varepsilon}{k_{\rm B} T}
\right)}
\end{equation}
where $M_{\alpha}$ and $N_{\alpha}$ are the mass and the number density of
the alpha particle, and $\hbar$ and $k$ are Planck's and Boltzmann's
constant, respectively. The temperature of the stellar plasma
is given by $T$.

The two quantities in Eq.~\ref{alphaa} that change their value by varying
the strength of the N-N interaction are $\varepsilon$ and $\Gamma_\gamma$
of the $0^+_2$ resonance in $^{12}$C. These quantities are calculated
in a microscopic 12-body,
three-alpha cluster model of $^{12}$C. The cluster model assumes relatively
simple harmonic-os\-cil\-lator shell-model states for the alpha particles and
describes with high precision the relative motions between the clusters, which
are the most
important degrees of freedom {\it (8)\/}. The model treats
$^{12}$C as a system of twelve interacting nucleons, properly takes into
account
the Pauli principle, and is free of any nonphysical center-of-mass excitation.
The only input parameter in the model is the effective N-N
interaction. To explore and understand any dependence of the results
on the chosen interaction, we used several different sets of
forces, including the Minnesota (MN) and modified Hasegawa-Nagagta (MHN)
interactions {\it (9-12)\/}. These interactions give a good overall
description of the $\alpha+\alpha$ scattering and the $^{12}$C levels
{\it (8)\/}.

As a first step of our calculations, we fine-tuned each force (by slightly
changing their exchange-mixture parameters) to fix $\varepsilon$ in
Eq.~\ref{alphaa} at its experimental value. Then we
performed calculations for $\varepsilon$ and $\Gamma_\gamma$ while slightly
varying the strengths of the N-N forces. For a given set of forces, all
repulsive and attractive terms of the interaction are multiplied by a factor
$p$, which was set between 0.994 and 1.006. Thus, the
calculations with $p=1.0$ correspond to the physical strength of the N-N
interaction and reproduce the experimental value of $\varepsilon$.

We find that the value of $\Gamma_\gamma$ is minimally changed by the small
variations of the N-N interactions, leading to negligible changes in
$r_{3\alpha}$. Thus, in the stellar model calculations we can fix
$\Gamma_\gamma$ to its experimental value in all cases. The resonance energy
$\varepsilon$, however, is rather sensitive to variations in the N-N force,
leading to large changes in the triple-alpha rate $r_{3\alpha}$.

In addition to the strong interaction, the triple-alpha reaction rate depends
on another fundamental force, the electromagnetic interaction. The strength
of the Coulomb interaction between the protons is proportional to the fine
structure constant $\alpha_{\rm f}$. To see the sensitivity of
$r_{3\alpha}$ to changes in $\alpha_{\rm f}$, we performed calculations for
the
energy
of the $0^+_2$ resonance of $^{12}$C while slightly varying $\alpha_{\rm f}$
from its experimental value (1/137.036),
much the same way as with the strong N-N
interaction. We found that varying $\alpha_{\rm f}$ leads to a smaller and
reversed effect for the change in the energy of the $0^+_2$ resonance.
This result is logical because the Coulomb interaction between the
alpha clusters has a different sign and is weaker than the strong
interaction.

The $^{16}{\rm O}(\alpha,\gamma){^{20}{\rm Ne}}$ reaction is nonresonant, so
variations in the strengths of the strong or Coulomb forces can have only small
effects on its reaction rate. The $^{12}{\rm C}(\alpha,\gamma){^{16}{\rm O}}$
process may look more dangerous because its cross section is strongly
affected by subthreshold states in the $^{16}$O nucleus {\it (5)\/}.
However, if the
N-N force is made weaker, then the subthreshold states become less bound,
thereby enhancing the $^{12}{\rm C}(\alpha,\gamma){^{16}{\rm O}}$
cross section.
Therefore, in the case of a weaker force, the small C/O ratio is still
decreased. An analogous reasoning holds for a stronger force. Thus, without
doing any calculation for the  $^{12}{\rm C}(\alpha,\gamma){^{16}{\rm
O}}$ and $^{16}{\rm O}(\alpha,\gamma){^{20}{\rm Ne}}$ reactions with the
modified forces, we can conclude that their effect would strengthen our
hypothesis regarding carbon and oxygen production.

For stars with masses more than $\sim$0.9 solar mass ($M_\odot$) the
temperatures in the center can reach the burning temperature of He, thus
producing C by the triple-alpha process {\it (13)\/}. Depending on the
exact
conditions in the center, some C may be further fused to
O by capturing an additional alpha particle. When all He in the
center of the star is consumed, the triple-alpha reaction proceeds in
shells around this core. In this phase, two burning shells are present,
one shell that burns H and one that burns He, and are separated by a region of
radiative energy transport.

In stars with masses $\lessim$$8M_\odot$ this \mbox{double} shell-burning
structure is thermally unstable {\it (14)\/}, creating thermal
pulses with high rates of mass loss. In this phase these
stars lose the whole envelope, i.e., the layers above the
H-burning shell, creating planetary nebulae
(PNe) with a white dwarf in the center. During one thermal pulse, the
H- and He-burning shells alternately provide the energy source of the
star.
When the energy-release of the He shell approaches its maximum, a
convective zone develops. The C and O produced by fusion of He
is dredged into layers that might, in a latter part of the pulse, be
covered by the convective envelope {\it (15)\/}. In this way, they are
transported to the
surface where they are blown away by the stellar wind, creating a
heavy-element enriched PNe.

Unfortunately, the exact amount of C and O dredged up is
uncertain, because it depends on poorly understood processes at
the boundary of the convective layers, such as overshooting or rotational
mixing {\it (16)\/}. Thus, theoretical predictions about the
composition of PNe are not accurate. But here we
were only interested in the change of C and O abundances under
modifications of the $0^+_2$ resonance energy in $^{12}$C. Because we did not
expect that additional mixing processes at the formal boundaries of
convective zones are altered by changes in nuclear physics, we
believed the relative
changes of the C and O abundances would remain independent of
such mixing processes.

In massive stars ($M\!>8\,M_\odot$) advanced burning phases take place
until the center reaches nuclear matter
densities, where the collapsing matter bounces back and explodes in
supernovae (SNe). Although further nuclear reactions take place in the
explosion, the relative
abundances of C and O are not believed to change drastically.

The composition of the interstellar material (ISM) is a mixture of ejecta from
stars with different masses. To determine which stellar mass
contributes to the total amount of C and O in the ISM, one has to
know these abundances in the PNe or SNe and the mass function, which provides
the number of stars with a certain mass. At present, it is unclear which
stars most enrich the ISM with C and O. Various calculations of stellar
yields of massive stars exist {\it (17-19)\/}, which agree
that O seems to be dominantly produced in these stars, whereas
the C situation is less clear. Recent investigations appear to
support the assessment that stars with masses of
$1-8\,M_\odot$ are the major producers of C {\it (20, 21)\/}.
\begin{figure*}[t]
\hbox to\hsize{\hss\epsfxsize=\textwidth\epsfbox{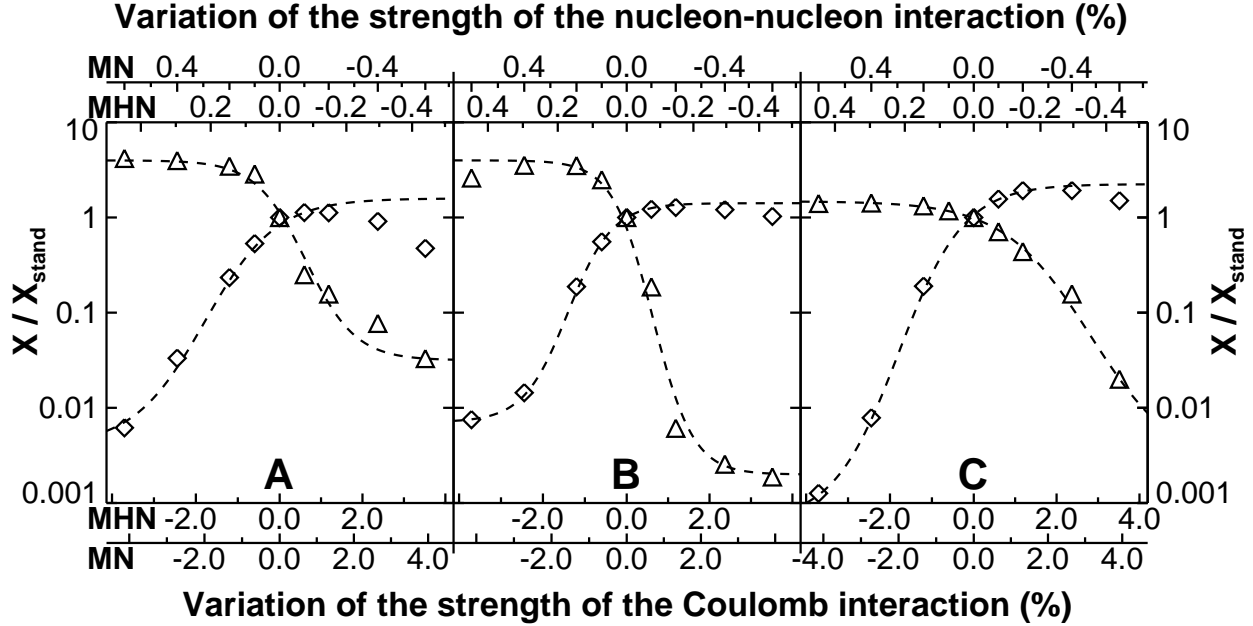}\hss}
\caption{\small The change of the C and O mass
abundances through variations of
the strengths of the strong and Coulomb interactions. The C
mass abundance, $X_{\rm C}\,(\triangle)$, and the O mass abundance,
$X_{\rm O}\,(\Diamond)$ are shown in (A), (B), and (C) for stars with masses
20, 5, and $1.3\,M_\odot$, respectively, in units of the standard values,
$X_{\rm stand}$.
The variations of the strength of the strong and
Coulomb interaction are given in the upper and lower scales for the two
effective N-N forces MHN and MN.
Dashed curves are drawn to
show the trends.}
\label{fig1}
\end{figure*}

The calculations with the reaction rates from the slightly modified strong
or Coulomb interaction were performed with a contemporary stellar evolution
code, which contains the most recent physics input {\it (22)\/}. In
particular, use of this code can produce up-to-date solar models {\it (23)\/}
and can allow one to follow the evolution
of low-mass stars through the thermal pulses of asymptotic
giant branch (AGB) stars {\it (24)\/}. The nuclear network is designed
preferentially to calculate the H- and He-burning phases in
low-mass stars. Additionally, the basic reactions of C- and
O-burning are included, which may destroy the produced C and O
in massive and intermediate-mass stars.

We performed stellar model calculations for a typical massive,
in\-ter\-me\-di\-a\-te-mass, and low-mass star with masses
20, 5, and 1.3\,$M_\odot$, respectively.
The stars are followed from the onset of H-burning until the third
thermal pulse in the AGB, or until the core temperature reaches $10^9$\,K in
the case of the $20\,M_{\odot}$ star (the nuclear network is not sufficient
to go beyond this phase). For the $1.3\,M_\odot$ star, which loses its
envelope by stellar winds during the thermal-pulse phase, the
maximum C and O abundances in the He-burning region have been extracted.
By taking the maximum abundances in this region, we have a measure of how
much the envelope of the star can be enriched by C or O,
irrespective of how efficient the dredge-up of heavy elements is
compared with our model.

For the three stellar masses, the evolution is calculated
with different values of the resonance energy in the triple-alpha reaction
within a range to cover variations in the strength of the strong and
Coulomb interaction up to 0.5 and 4\,\%, respectively. The
resulting modifications in the C and O abundances are shown
(Fig.~1) with
respect to the case, where the standard value of the resonance energy has
been used (i.e., with no variations of the strength of the strong or Coulomb
interaction). Because each shift in the resonance energy can be
identified with a
variation in the strength of the N-N or Coulomb interaction, we
scaled the upper and lower ordinate with variations in these
quantities.
Our calculations indicate that the behavior of the
residual alpha-alpha interaction, and thus that of the resonance energy
of the $0_2^+$ state, is expected to lie somewhere between the predictions
of two of our effective N-N interactions, the MN {\it (10, 11)\/}
and the MHN {\it (12)\/} forces.
Therefore, we show the abundances calculated only with these
two effective N-N interactions (Fig.~1).

A~saturation~of~the~C~production is reached
with~increasing~N-N~\mbox{interaction}~(very~pro\-
nounced~for~the~$5\,M_{\odot}$~star)
because no alpha particles are available below the
He-burning front. Thus, the star does not gain additional energy from
the $^{12}{\rm C}(\alpha,\gamma){^{16}{\rm O}}$ reaction. The stellar core
contracts more rapidly and C-destroying $^{12}{\rm C}+{^{12}{\rm C}}$
reactions ignite earlier. For O, a similar behavior can be observed with
decreasing N-N interaction strength. Because the temperatures where the
triple-alpha reactions set in are enhanced with decreasing N-N force, the
temperatures below the He-burning shell are much higher than in the
standard case, and $^{16}{\rm O}+\alpha$ reactions can destroy the
previously generated O more efficiently than in the standard case
(0\,\%).

We conclude that a change of more than 0.5\,\% in the strength of
the strong interaction or more than 4\,\% change
in the strength of the Coulomb force would destroy either nearly all C or
all O in every star. This implies that irrespective of
stellar evolution the contribution of each star to the abundance
of C or O in the ISM would be negligible. Therefore, for the
above cases the creation of
carbon-based life in our universe would be strongly disfavoured. The
anthropically allowed strengths of the strong and electromagnetic forces
also constrain the Higgs vacuum expectation value {\it (25)\/} and yield
tighter
constraint on the quark masses than do the constraints from light nuclei {\it
(26)\/}.
Therefore, the results of this work are relevant not only for the anthropic
cosmological principle {\it (27)\/}, but also for the mathematical design of
fundamental
elementary particle theories.

\newpage
{\large \bfseries{References and Notes}}

\small
\begin{enumerate}
\setlength{\parsep}{0ex}\setlength{\itemsep}{-.5ex}
\item D.N.F.~Dunbar, R.E.~Pixley, W.A.~Wenzel, W.~Whaling, {\it
Phys. Rev.}~{\bf 92}, 649 (1953).
\item F.~Hoyle, D.N.F.~Dunbar, W.A. Wenzel, {\it Phys.~Rev.}~{\bf 92},
1095 (1953).
\item C.W.~Cook, W.A.~Fowler, T.~Lauritsen,
{\it Phys.~Rev.}~{\bf 107}, 508 (1957).
\item V.S.~Shirley, Ed.,{\it~Table of Isotopes,
Volume I\/} (John Wiley \& Sons, New York, ed.~8, 1996).
\item C.E.~Rolfs, W.S.~Rodney, {\it Cauldrons in the Cosmos\/}
(Univ.~of Chicago Press, Chicago, 1988).
\item M.~Livio, D.~Hollowell, A.~Weiss, J.W.~Truran,
{\it Nature\/} {\bf 340}, 281 (1989).
\item H.~Oberhummer, A.~Cs\'ot\'o, H.~Schlattl,
Pre-print available at \\ http://xxx.lanl.gov/abs/astro-ph/9908247.
\item R.~Pichler, H.~Oberhummer, A.~Cs\'ot\'o,
S.A. Moszkowski, {\it Nucl.~Phys.~A\/} {\bf 618}, 55 (1997).
\item A.B.~Volkov,
{\it Nucl.~Phys.\/} {\bf 74}, 33 (1965).
\item I.~Reichstein, Y.C.~Tang,
{\it Nucl.~Phys.~A\/} {\bf 158}, 529 (1970).
\item D.R.~Thompson, M.~LeMere, Y.C.~Tang,
{\it Nucl.~Phys.~A\/} {\bf 286}, 53 (1977).
\item H.~Furutani {\it et al.\/},
{\it Prog.~Theor.~Phys.~Suppl.\/} {\bf 68}, 193 (1980).
\item R.~Kippenhahn, A.~Weigert, {\it Stellar Structure and
Evolution\/} (Springer-Verlag, Berlin, 1990).
\item M.~Schwarzschild, A.R.~H\"{a}rm,
{\it Astrophys.~J.\/} {\bf 142}, 855 (1965).
\item I.~Iben, A.~Renzini,
{\it Ann.~Rev.~Astron.~Astrophys.\/} {\bf 21}, 271 (1983).
\item A.~Weiss, P.A.~Denissenkov, C.~Charbonnel,
{\it Astron.~Astrophys.\/} {\bf 356}, 181 (2000).
\item
S.E.~Woosley, T.A.~Weaver,
{\it Astrophys.~J. Suppl.~Ser.\/} {\bf 101}, 181 (1995).
\item
K.~Nomoto {\it et al.\/},
{\it Nucl.~Phys.~A\/} {\bf 616}, 79 (1997).
\item
A.~Maeder, {\it Astron.~Astrophys.\/} {\bf 264}, 105 (1992).
\item F.X.~Timmes, S.E.~Woos\-ley, T.A.~Weaver, {\it
Astrophys.~J.~Suppl.~Ser.\/} {\bf
98}, 617 (1995).
\item R.B.C.~Henry, K.B.~Kwitter, J.A.~Bates, {\it Astrophys.~J.\/}, in press.
Pre-print available at http://xxx.lanl.gov/abs/astro-ph/9910347.
\item
A.~Weiss, H.~Schlattl, {\it Astron.~Astrophys. Suppl.} {\bf 144}, 487 (2000).
\item
H.~Schlattl, A.~Weiss,
{\it Astron.~Astrophys.\/} {\bf 347}, 272 (1999).
\item
J.~Wagenhuber, A.~Weiss,
{\it Astron.~Astrophys.\/} {\bf 286}, 121(1994).
\item E.T.~Jeltema, M.~Sher,
{\it Phys.~Rev.~D\/} {\bf 61}, 017301 (2000).
\item C.J.~Hogan,
Pre-print available at \\ http://xxx.lanl.gov/abs/astro-ph/9909295.
\item J.D.~Barrow,~F.J.~Tipler,~{\it The~Anthropic} {\it Cosmological~
Principle}~(Oxford University Press, Oxford, 1986).
\item Supported~in~part~by~the~Fonds~zur wis\-senschaftlichen
Forschung~in~\"Os\-ter\-reich (P10361-PHY), the OTKA Fund (D32513), the
Education Ministry
(FKFP-0242/2000), the Academy (BO/00520/98) in Hungary, and by the John
Templeton Foundation (938-COS153).
\end{enumerate}
\end{document}